\documentclass[twocolumn,aps,epsf,floatfix,showpacs]{revtex4}

\newcommand{\ket}[1]{\left |#1\right >}

\newcommand{\expec}[1]{\left < #1\right >}
\usepackage{amsmath}
\usepackage{amsfonts}
\usepackage{graphicx}
\usepackage{amsthm}
\usepackage{subfigure}

\usepackage{hyperref}

\renewcommand{\v}[1]{\boldsymbol{#1}}

\newcommand{\adi}{a_{i}^{\dagger} }
\newcommand{\ai}{{a}_{i} }

\newcommand{\aj}{{a}_{j} }

\newcommand{\bdi}{b_{i}^{\dagger} }

\newcommand{\bj}{{b}_{j}}

\newcommand{\lp}{\left ( }
\newcommand{\rp}{\right ) }
\newcommand{\defn}{\buildrel \text{def.} \over \equiv}
\newcommand{\hc}{\text{H.c.}}

\newcommand{\mc}{\mathcal}

\newcommand{\beq}{\begin{eqnarray*}}
\newcommand{\eeq}{\end{eqnarray*}}

\newcommand{\be}{\begin{eqnarray}}
\newcommand{\ee}{\end{eqnarray}}

\def\lsim{\mathrel{\rlap{\lower4pt\hbox{\hskip1pt$\sim$}}
    \raise1pt\hbox{$<$}}}                

\def\gsim{\mathrel{\rlap{\lower4pt\hbox{\hskip1pt$\sim$}}
    \raise1pt\hbox{$>$}}}                

\begin{document}


\bibliographystyle{prsty}

\title{
Hyperfine spectra of trapped Bosons in optical lattices}
\author{Kaden R.A. Hazzard} \email{kh279@cornell.edu}
\affiliation{Laboratory of Atomic
and Solid State Physics, Cornell University, Ithaca, New York 14853}

\author{Erich J. Mueller}
\affiliation{Laboratory of Atomic and Solid State Physics, Cornell
University, Ithaca, New York 14853}

\begin{abstract}
We calculate the
interaction induced inhomogeneous broadening of spectral lines in a trapped Bose gas as a function of the depth of a three-dimensional cubic optical lattice.
  As observed in recent experiments, we find that the terraced ``wedding-cake" structure of Mott plateaus
  splits the spectrum into a series of discrete peaks.
The spectra are extremely sensitive to density corrugations and trap anharmonicities.
For example, even when the majority of the cloud is superfluid the spectrum displays discrete peaks.
\end{abstract}

\maketitle

\section{Introduction}\label{introduction}

The study of quantum degenerate atoms confined to periodic potentials forms an important subfield of modern atomic physics.
Research in this area is driven by its connection to
 condensed matter physics and
  quantum information processing~\cite{brennen:qc-init,rabl:qc-init}.
A rich set of probes, including
optical spectroscopy, noise spectroscopy, interference, and
density profile measurements~\cite{greiner:mott-orig,foelling:noise,foelling:realspace,greiner:freeze-pops,
bloch:review,campbell:ketterle-clock-shift}, have been used to characterize these systems, with a focus on understanding the interaction driven superfluid-insulator transition.
  Here we analyze in detail what information one gains from inhomogeneous pressure shifts of spectral lines in a gas of bosons confined to an optical lattice.

Atomic interactions lead to pressure and density  dependent shifts of atomic lines.  These ``clock shifts" limit the accuracy of atomic clocks.
In an inhomogeneous system they are spatially dependent, yielding a broadened spectrum whose structure reveals details about the local atomic correlations.  In many situations the clock shift is proportional to the atomic density, and the spectral line directly gives a histogram of the atomic density.  As an example of this technique, Bose-Einstein condensation in spin polarized atomic hydrogen was detected through the line shape of a two-photon 1s-2s transition~\cite{fried:hydrogen}.
More recently, Campbell {\em et al.}~\cite{campbell:ketterle-clock-shift}
utilized atomic
clock shifts to experimentally probe bosons trapped in an optical lattice, finding evidence for Mott insulating shells.
Motivated by these latter experiments, we present a theoretical analysis of the lineshapes which should
be found when bosonic atoms in a periodic potential are confined by a nominally harmonic potential.

\begin{figure}
\setlength{\unitlength}{1.0in}
\centering
\hspace{-0.05in}
\includegraphics[width=3.25in,angle=0]{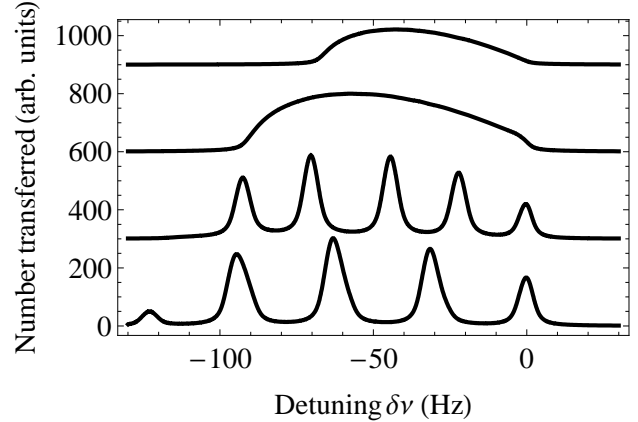}
\caption{ \label{fig:cs-gen} Theoretical spectra 
showing the number of $^{87}$Rb atoms transferred from hyperfine state $|a\rangle=\ket{F=1,m_F=-1}$ to state $|b\rangle=\ket{F=2,m_F=1}$ when excited by light detuned from resonance by the frequency $\delta\nu$.  The $N=9\times 10^4$ atoms are confined by a harmonic potential with $\bar\omega=(\omega_x \omega_y\omega_z)^{1/3}=2\pi \times100$Hz and a three-dimensional periodic potential with lattice depth
$V_0=5, $ 10,  25, $35E_{\text{rec}}$ (from top to bottom).
}
\end{figure}

In Sec.~\ref{cs-basics} we use a local density approximation to calculate the spectrum of a harmonically trapped gas as a function of the depth of an optical lattice (Fig.~\ref{fig:cs-gen}).  In the deep lattice limit,
 the spectral line splits into several distinct peaks, associated with the formation of density plateaus.  Due to the sensitivity of these spectra to small density corrugations, this splitting occurs even when large sections of the cloud are superfluid.  Despite  qualitative agreement with experiments, our calculation severely underestimates the  small detuning spectral weight.
In Sec.~\ref{cs-refinements} we show that these discrepancies are consistent with
 trap anharmonicities.  We also explore other sources for the discrepancy.

\textbf{Experimental Details.}  Since we are largely concerned with the experiment in Ref.~\cite{campbell:ketterle-clock-shift}, we give a brief review of the important experimental details.  In these experiments,
a gas of $^{87}$Rb
atoms in the $\ket{a}=\ket{F=1,m_F=-1}$ hyperfine state ($F$ the total spin
and $m_F$ its $z$ component) was cooled well below the
condensation temperature~\cite{kett-priv-comm}.
By combining optical and magnetic fields, a three-dimensional periodic potential $V_{\text{per}}=-V_0 \left[\cos(2 \pi x/d)+\cos(2 \pi y/d)+\cos(2 \pi z/d)\right]$
was superimposed on a trapping potential.
The spacing between
lattice sites, $d=\lambda/2=532$nm, is half  of the lattice lasers' wavelength.
The lattice depth $V_0$
was
tuned from zero to $40 E_{\text{rec}}$
where $E_{\text{rec}}=\frac{\hbar^2}{2m}\lp
\frac{2\pi}{\lambda}\rp^2$ is the photon recoil energy.
A microwave and RF field were tuned near resonance for a two photon   transition from the $\ket{a}$ state to an excited hyperfine state $\ket{b}=\ket{F=2,m_F=1}$.

\section{Spectrum of harmonically trapped gas\label{cs-basics}}

\subsection{Hamiltonian and approximations}

\textbf{Hamiltonian.} Bosons in a sufficiently deep optical
lattice are described by the Bose-Hubbard model~\cite{jaksch:olatt}, found by projecting the
full Hamiltonian onto the lowest Bloch band.
We will work with a two-internal state Bose-Hubbard Hamiltonian, where $a_i$ and $b_i$ annihilate bosons at site $i$ in states $\ket{a}$ and $\ket{b}$, respectively.  Including an external trapping potential the Hamiltonian is
\begin{eqnarray}
H &=& -t_a\sum_{<i,j>} \adi \aj + \sum_i
\left[\frac{U_{a}}{2}n_{i,a}(n_{i,a}-1)  +V_{i,a} n_{i,a} \right] \nonumber\\
    && {}-t_b \sum_{<i,j>} \bdi \bj + \sum_i
    \left[\frac{U_{b}}{2}n_{i,b}(n_{i,b}-1) +V_{i,b} n_{i,b}\right ]\nonumber\\
    && {}+ U_{ab}\sum_i  n_{i,a} n_{i,b} +\hc \label{Ham-both}
\end{eqnarray}
where $n_{i,\alpha}\defn \alpha^\dagger_i \alpha_i$.
The $t_\alpha$'s describe hopping rates
and $U_{\alpha \beta}$ the interaction where $\alpha$ and $\beta$ label the species ($\ket{a}$ or $\ket{b}$).
We abbreviate $U_{\alpha}=U_{\alpha\alpha}$.  We have absorbed the chemical potentials into the trapping potential, writing $V_{i,\alpha}=\mc V_{i,\alpha}-\mu_\alpha$ where $\mc V_{i,\alpha}$ is the external potential at site $i$ for species $\alpha$.
In terms of microscopic quantities, these parameters are
$t_\alpha = \int d\v{r} \, w_\alpha^*(\v{r}) \left[ -\hbar^2/(2m_\alpha) \nabla^2+V_{\text{per}}(\v{r})\right] w_\alpha(\v{r}) $, $
\mc V_{i,\alpha}
\approx V_{\text{trap}}(\v{R_i}) $, and
$U_{\alpha\beta} = \lp4\pi\hbar^2a_{\alpha \beta}/m\rp$ $\int d\v{r}$ $|w_\alpha(\v{r})|^2|w_\beta(\v{r})|^2$
where $m_\alpha$ is the mass and  $w_\alpha$ the normalized Wannier function for state $\alpha$, while  $a_{\alpha \beta}$ denotes the $\alpha$-$\beta$ scattering length.
For $^{87}$Rb, the relevant scattering lengths are $a_{aa}=5.32$nm and $a_{ab}$=5.19nm~\cite{kempen:scatlen}. The $\ket{b}$ atoms will be sufficiently dilute that $a_{bb}$ will not enter our calculation.  The competition
between the  kinetic and interaction  terms drives the Mott insulator to superfluid phase transition.

In the experiments of interest, the atoms all begin in the $\ket{a}$ state, and one measures the rate at which atoms are transferred to the $\ket{b}$ state under the influence of a weak probe of the form $H_{\rm probe}\propto\sum_j b_j^\dagger a_j e^{-i \omega t}+{\rm H. c.}$, within the rotating wave approximation, where $\omega$ is the frequency of the photons.  To calculate this response, it is sufficient to understand the properties of the single-component Bose-Hubbard model (the terms in Eq.~(\ref{Ham-both}) containing only $a$'s).

\textbf{Mean-field theory.}
The   ground state of the single component Bose-Hubbard model is well approximated by the Gutzwiller mean-field theory (GMF) of Ref~\cite{fisher:bhubb}.  This approach is exact in infinite dimensions and in the deep Mott insulator and superfluid limits.  Sophisticated numerical
calculations, some with a trapping potential,
have shown that this mean field theory yields qualitatively accurate phase diagrams, energies,
and spatial density profiles~\cite{pollet:mi,bergkvist:mi,wessel:mi,batrouni:mi,demarco:stability}.
As a point of reference, Monte-Carlo calculations predict that for unity filling the 3D Bose-Hubbard model on a cubic lattice has an insulator-superfluid transition at $t/U=0.03408(2)$, while mean field theory gives $t/U=0.029$.  We will work within this approximation.
As will be apparent, one could extend our results to include fluctuation effects by numerically calculating the density and compressibility of the homogeneous system.  Within the local density approximation discussed below these homogeneous quantities are the only theoretical input needed to determine the spectrum.

The Gutzwiller mean field approximation to the Bose-Hubbard model can be developed either from a mean-field or variational
standpoint.
As a variational ansatz, GMF corresponds to taking the
wavefunction to be a tensor product of states at each site:  $\ket{\Psi}=\bigotimes_i \lp\sum_n f^{(i)}_n\ket{i,n}\rp$
where   $\ket{i,n}$ is the state with $n$
particles at the $i$'th site; the $f_n^{(i)}$ are varied.
In  the corresponding mean-field language, fluctuations of the annihilation operators
from their expectations are assumed not to affect neighboring sites. Then,
assuming
translation symmetry  remains unbroken and letting $q$ be the number of nearest neighbors, one has
\begin{eqnarray}
H_{\text{MF}}
    &=&\sum_{i}\left[-q t \adi
\langle a\rangle +
U\frac{n_i^2}{2} +V_i n_i +\hc\right] \label{Ham-MF}
\end{eqnarray}
from which one can find a self-consistent ground state with  $\expec{a}=\sum_n \sqrt{n+1} f^*_n f_{n+1}$.

\textbf{Local density approximation.}
We use a local density (or Thomas-Fermi) approximation (LDA) to calculate the spatial
dependence of thermodynamic quantities:
all physical quantities at location $\v{r}$ are taken to be those of a homogeneous system at a chemical potential $\mu-\mc V(\v{r})$.  This is expected to be valid when the spatial correlation length of the  homogeneous system is much smaller than the length scale of the trapping potential~\cite{pethick:p-s-book}.
The validity of the GMF+LDA is discussed in Ref.'s~\cite{pollet:mi,bergkvist:mi,wessel:mi,batrouni:mi,demarco:stability}.

\subsection{Homogeneous clock shifts}

The  clock shift is
a density-dependent shift in the energy splitting $\Delta$ for driving a transition from  internal atomic states $\ket{a}$ to states $\ket{b}$ due to the
inter-particle interactions.  In this section we review the known results for the clock shift of a homogeneous system in terms of local correlations, and specialize to the case of atoms in a periodic potential.

We will assume that $t_a=t_b$ and $V_{i,b}=V_{i,a}+\Delta_0$ where $\Delta_0$ is the
energy splitting of the two states in vacuum.
 These assumptions are justified in the recent experiments, where the polarizabilities of the two internal states are nearly indistinguishable.
In linear response, the average clock shift energy of the homogeneous system is then
\be
\delta E_2
    &\doteq& \lp U_{ab}-U_{\alpha}\rp\frac{\expec{\sum_i \adi \adi \ai \ai}}{\expec{\sum_i \adi \ai}} \label{2-body-shift}
\ee
where the expectation is in the initial, all-$\ket{a}$ state~\cite{oktel:cs-ref2,pethick:pseudopot-breakdown}.  This expression can be rewritten in  a somewhat more familiar form by defining the local second order correlation function $g_2 \defn \expec{\adi \adi \ai \ai}/\expec{\adi \ai}^2$ so
\be
\delta E_2     &\doteq& \lp U_{ab}-U_{\alpha}\rp g_2\expec{n}.\label{2-body-shift-o-l}
\ee

\textbf{Special cases of the clock-shift formula: dilute superfluid, Mott insulator, and normal fluid. }
For a dilute  superfluid, the initial state is a coherent state, and Eq.~(\ref{2-body-shift}) gives a shift proportional to the occupation of each site,
\be
\delta E_{\text{SF}} &=& (U_{ab}-U_{a})n.\label{SF-cs}
\ee
Deep within the Mott insulating phase, the initial state is a number eigenstate and
\beq
\delta E_{\text{MI}} &=&(U_{ab}-U_{a})(n-1).
\eeq
This latter formula has an intuitive explanation.  In a Mott insulator with  filling of one particle per site, the atoms are isolated so  there is no interaction between particles.
Hence $\delta E_{\text{MI}}$ must vanish when $n=1$.
Fig.~\ref{fig:cs-full} illustrates how the clock shift energy evolves from being proportional to $n$ to $n-1$ by juxtaposing the contours of fixed $\delta E_2$ and those of fixed density.

If one raises the temperature the system becomes a normal fluid, even at weak interactions.  In the absence of interactions, the normal fluid statistical factor $g_2$ appearing in Eq.~(\ref{2-body-shift-o-l}) is
$g_2 = 2$,~\cite{landau:ll-sm} so that the clock shift energy is twice as large as in the superfluid:
\be
\delta E_{\text{NF}} &=& 2(U_{ab}-U_a) n. \label{nf-cs}
\ee  Given that there is no phase transition between the zero temperature Mott insulator and the normal gas, it is interesting that the clock shift energy changes from $2(U_{ab}-U_a)n$ in the normal fluid to  $(U_{ab}-U_a)(n-1)$ in the Mott insulator.  A quantitative understanding of this crossover would require calculating the temperature dependence of the pair correlations in the strongly interacting limit.

\subsection{Calculation of spectrum in a trap}

To calculate the spectrum we assume that the gas can be treated as locally homogeneous, and we can independently sum the spectrum from each region in the cloud.  As in the experiment, we imagine applying a weak probe at frequency $\omega$ for a time $\tau\approx 100$ms.  Assuming that the finite probe duration is the primary source of broadening, second order perturbation theory implies that
the number of atoms of atoms transferred to the $\ket{b}$ state will be
\begin{equation}\label{csb}
N_b(\omega)\propto \int d^3 r\, n(r) \delta_{1/\tau}(\Delta(\v{r})-\omega),
\end{equation}
where $\delta_\gamma(\nu)$ has a peak of width $\gamma$ at $\nu=0$.  We will model
$\delta_\gamma(\nu)=(1/\pi)\gamma/(\nu^2+\gamma^2)$ as a Lorentzian of width $\gamma$.
 The local density $n(r)$ and clock shift $\Delta(r)$ are set equal to that of a homogeneous system with chemical potential $\mu(\v{r})=\mu_0-V_{\text{trap}}(\v{r})$.
 Experimentally, the number of $\ket{b}$-atoms is measured by monitoring the absorption of a laser which transfers $\ket{b}$ atoms into a third state.

\begin{figure}[tbph]
\setlength{\unitlength}{1.0in}
\centering
\mbox{
\hspace{-0.15in}
    \subfigure[]
{
\includegraphics[width=1.65in,angle=0]{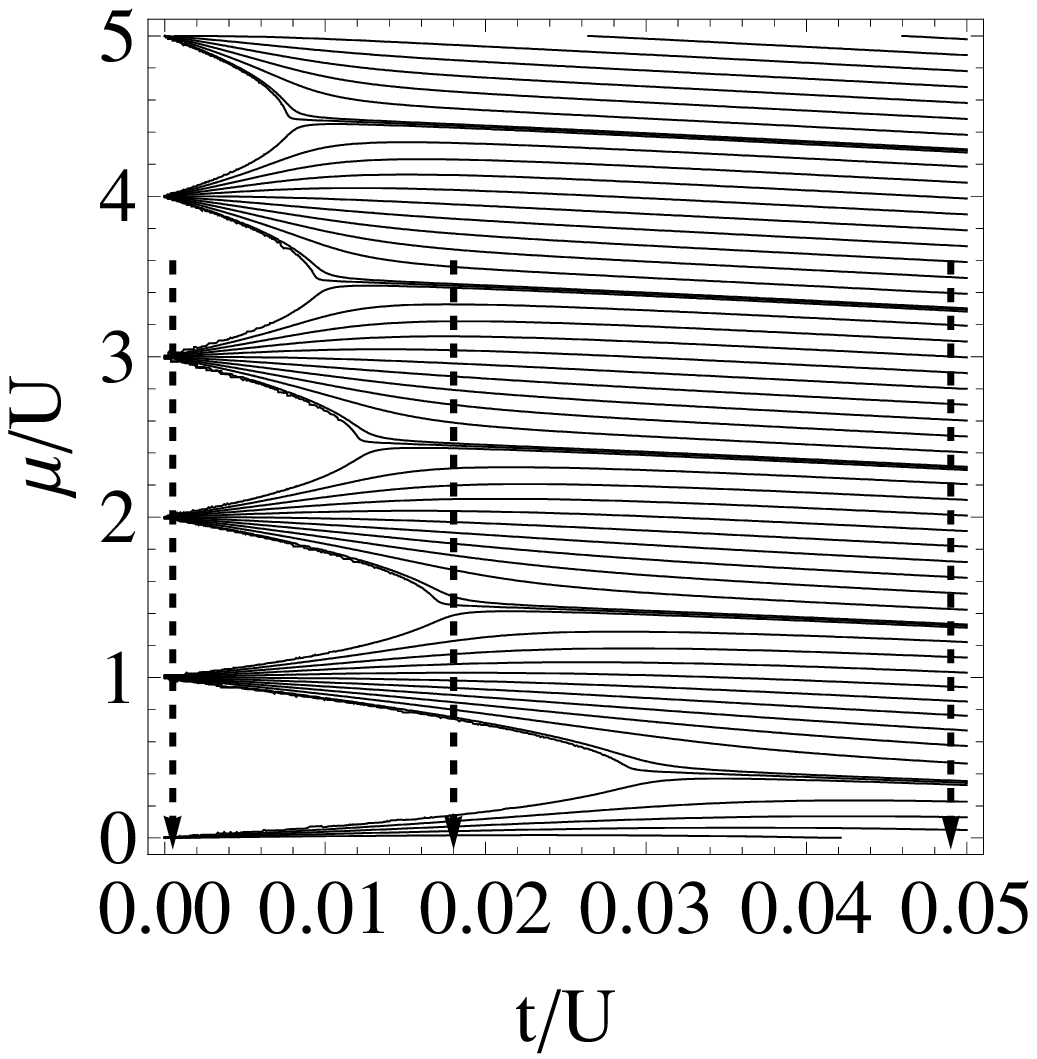}
\label{fig:GMF-LDA-phase-diagram}
}
\subfigure[]
{
\includegraphics[width=1.65in,angle=0]{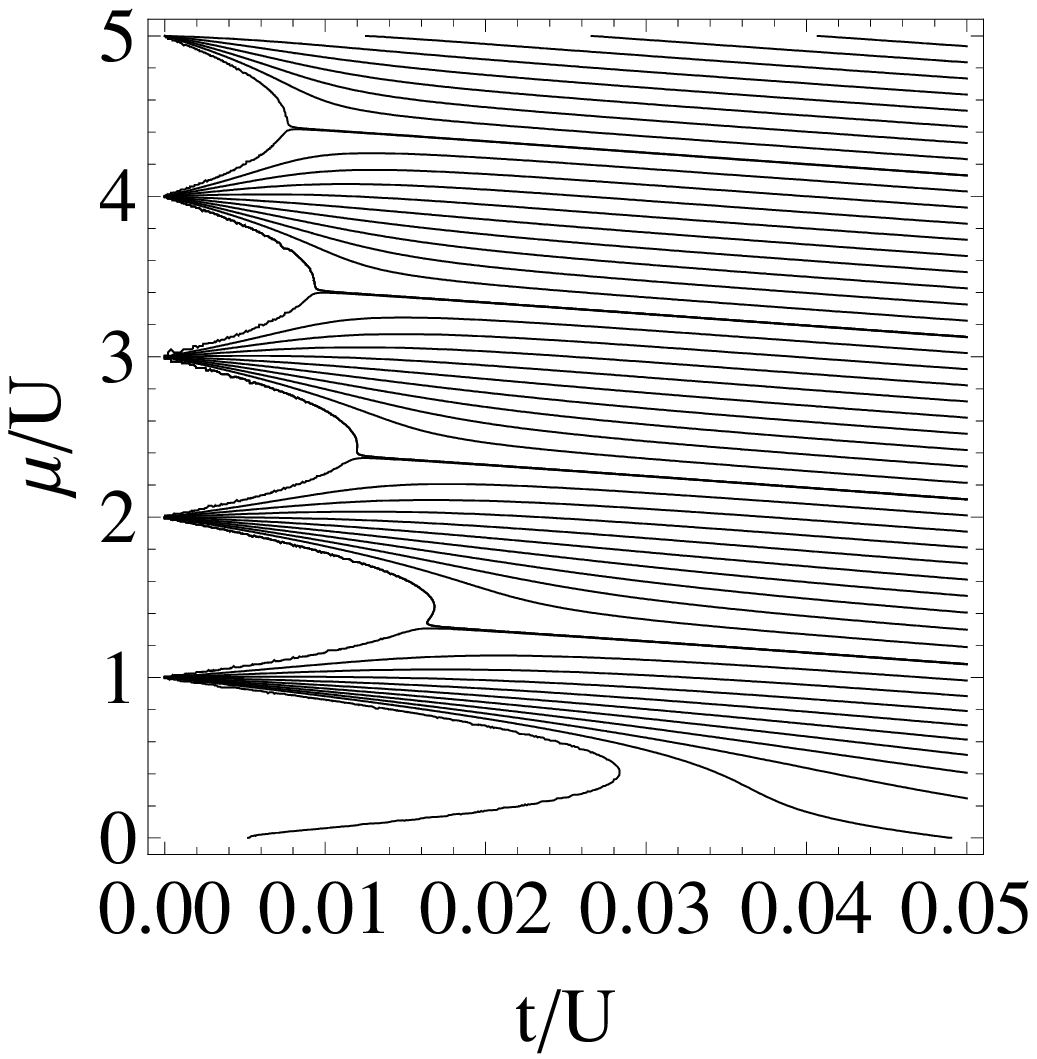}
\label{fig:cs-propto-dens}
}
}
\caption{ (a) Gutzwiller Mean Field phase diagram with constant density contours. The vertical dashed lines show the spatial dependance of the chemical potential for a trapped gas in the LDA: from left-to-right these correspond to  the deep Mott limit, the ``corrugated superfluid'' situation appropriate to Fig.~\ref{fig:cs-warning-corrug}, and the dilute superfluid.  (b) Phase diagram with contours of constant $\langle n \rangle g_2=\expec{n(n-1)}/\expec{n}$.   Contours are spaced by $0.1$, with additional lines at $m\pm0.01$, for integer $m$, to emphasize the Mott regions.
}\label{fig:cs-full}
\end{figure}

We calculate the integral in Eq.~(\ref{csb}) within the Gutzwiller mean field approximation to the Bose-Hubbard model.  As illustrated in Fig.~\ref{fig:cs-full},  both the density $n$ and the clock shift $\Delta$ can be expressed as functions of the parameters $\mu/U$ and $t/U$.
Within the local density approximation, $t$ is constant throughout the trap, and $\mu$ varies in space, taking its maximal value $\mu_0$ at the center of the trap.
For extreme values of $t/U$ (either large or small) we can analytically calculate the contours in Fig.~\ref{fig:GMF-LDA-phase-diagram}.  Generically, however we must rely on numerical methods.

Our results are shown in Fig.~\ref{fig:cs-gen} for a harmonic trap $V_{\text{trap}}(\v{r})=m\omega_x^2 x^2/2+m\omega_y^2 y^2/2+m\omega_z^2 z^2/2 $.
One sees that in the deep Mott limit, the spectrum displays sharp peaks, while in the deep superfluid limit, the lineshape is smooth.  The peaks are due to the stepwise variation of $\Delta(\mu)$ (illustrated in
 Fig.~\ref{fig:GMF-LDA-phase-diagram}) which lead to
large regions of the trap where $\Delta(r)$ takes on discrete values.   Compared with the experiments in Ref.~\cite{campbell:ketterle-clock-shift}, our spectral lines have severely reduced small detuning spectral weight.  In Sec.~\ref{cs-refinements} we show that trap anharmonicities can account for this difference.

Note that within the local density approximation, the spectrum is independent of trap anisotropies, as long as the trap is harmonic.  This generic feature of the LDA is seen by examining an integral of the form $I=\int d^3r\, f(\mu)$$=$$\int d^3r\,f(\mu_0-m\omega_x^2 x^2/2$$-$$ m\omega_y^2 y^2/2$$-$$m\omega_z^2 z^2/2)$.  Rescaling the coordinates so that $m\omega_x^2 x^2/2=\mu_0 \bar x^2$ (and similarly for $y$ and $z$), this integral becomes $I= \sqrt{8 \mu_0^3/m^3 \omega_x^2 \omega_y^2\omega_z^2} 4\pi \int d \bar r\,\bar r^2 f(\mu_0-\mu_0 \bar r^2)$, where $\bar r=\sqrt{\bar x^2+\bar y ^2+\bar z^2}$.  From this analysis it is clear that apart from an overall scale factor, the spectral lineshape is only a function of the central chemical potential $\mu_0$.

\textbf{Experimental Parameters.}
The experimental control parameters are the optical lattice depth $V_0$,  the number of particles $N$, and the trap frequencies $\omega_\nu$.  The natural theoretical parameters are $t, U,$ and $\mu_0$.
To compare our results to experiment, we use a non-interacting band structure calculation to relate $t$ and $U$ to $V_0$ \cite{jaksch:olatt}.  To relate $\mu_0$ to experimental parameters we note that within the LDA the number of trapped atoms $N$ is only a function of $t/U$, and $\mu_0/\hbar\bar\omega$ where $\bar\omega^3=\omega_x\omega_y\omega_z$.  For each value of $t/U$ we compute $N(\mu_0/\hbar\bar\omega)$, for several values of $\mu_0$, then invert the function to get $\mu_0$ as a function of $N$.
We then have the ability to select the value of $\mu_0$ corresponding to the number of particles used in the experiment.

Campbell {\em et al.}~\cite{campbell:ketterle-clock-shift} do not report the number of particles in the experiment.
For Fig.~\ref{fig:cs-gen}, we choose $N=9\times10^4$  so that the maximum site filling for $V_0=35 E_{\text{rec}}$ and $V_0=25 E_{\text{rec}}$ is $n=5$, as is observed in the experiment.

\subsection{Analytic Results}
\subsubsection{Dilute superfluid}

Having numerically calculated the spectra, we now
specialize to the dilute superfluid limit where the line shape can be calculated analytically.   The clock shift energy in this limit
is $\Delta=(U_{ab}-U_{a})\expec{n}$, and  within the local density approximation the site filling at position ${\bf r}$ is the greater of zero and
\be
n(r)  &=& \lp\mu_0 -\epsilon(0)-V_{\rm trap}({\bf r})\rp/U_{a}\label{T-F-bosons}
\ee
where as previously stated, $V_{\rm trap}({\bf r})=m\omega_x x^2/2+m\omega_y^2/2+m\omega_z z^2/2$ is the trapping potential, $\mu_0$ is the central chemical potential, and $\epsilon(0)$ is the energy of the $k=0$ single particle state in the periodic potential.  In the tight binding limit,  $\epsilon(0)=-q t$ where $q$ is the number of nearest neighbors.
Substituting this result into Eq.~(\ref{csb}), and neglecting the broadening one finds
\be
N_b(\delta \omega) &\propto&   \delta \omega \sqrt{ (U_{ab}-U_{a}) n(0)-\delta\omega },
\label{sf-clock-diag-2}
\ee
where $n(0)=(\mu_0-\epsilon(0))/U_a$ is the central density.
Similar expressions were obtained in Ref.~\cite{stenger:clock-shifts}

At fixed central chemical potential (equivalently, fixed central density) the width of the spectrum is proportional to $U_{ab}-U_a$. Given a fixed number of particles, the central density varies as $n(0)\sim 1/U^{3/5}$, so the width of the spectral line varies as $Un(0)\sim U^{2/5}$.

\subsubsection{Deep Mott limit}

Now we analytically calculate the spectrum in the deep Mott insulator limit, where the density of the homogeneous system with chemical potential $\mu$ equals the largest integer bounded by  $\mu/U$, denoted $\lceil \mu/U\rceil$~\cite{demarco:stability,jaksch:olatt}.
In the local density approximation the density jumps from density $n-1$ to $n$ as one moves through the location in space where  local chemical potential is given by $\tilde \mu=U_{a}(n-1)$.  Each plateau of fixed $n$ gives a (broadened) delta-function contribution to the line shape at detuning $\delta\omega_n=2 (n-1) (U_{ab}-U_a)$.  The magnitude of the delta function is proportional to the number of particles in the plateau, leading to a spectrum
\begin{eqnarray}\label{mi-clock-diag-1}
N_b(\delta\omega)&\propto&\sum_{n=1}^{\bar n} A_n \delta_{1/\tau}(\delta\omega-\delta \omega_n)\\\nonumber
A_{\bar n}&=& \left[{\mu_0-U_a ({\bar n}-1)}\right]^{3/2}  \bar n\\\nonumber
A_{n\neq \bar n}&=&  \left[\lp{\mu_0-U_{a}(n-1)}\rp^{3/2}-\lp{\mu_0-U_{a} n}\rp^{3/2}\right] n,
\end{eqnarray}
where $\bar n=\lceil \mu_0/U_a \rceil$ is the central density. 

\begin{figure}[tbph]
\setlength{\unitlength}{1.0in}
\centering
\includegraphics[width=3.25in,angle=0]{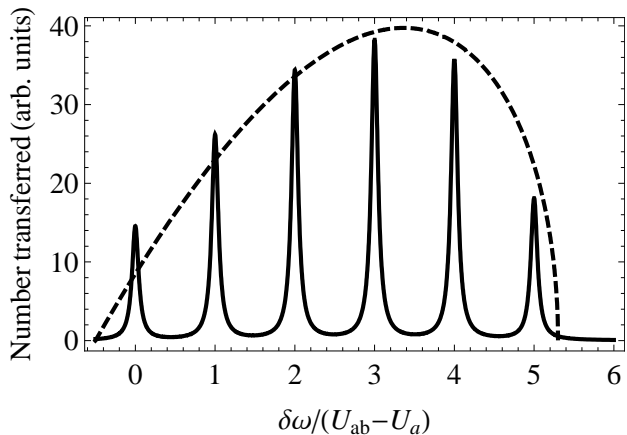}
\caption{
\label{fig:cs-deep-limits}
Analytically calculated spectra for the harmonically trapped system in the deep Mott limit (solid line), plotted as a function of $\delta \omega/(U_{ab}-U_a)$.  Superimposed is the spectrum of the superfluid (dashed line) with the same parameters, but horizontally shifted to the left by $-\delta \omega/2(U_{ab}-U_a)$.  The central density is $n_s(0)=[\mu_0-\epsilon(0)]/U_a=5.8$.  This illustrates that the envelope of the spectral line in the Mott insulating state has the same shape as the superfluid spectrum, shifted horizontally. }
\end{figure}

The deep superfluid and deep Mott insulator spectra are plotted in Fig.~\ref{fig:cs-deep-limits}
using Eq.'s~(\ref{sf-clock-diag-2}) and ~(\ref{mi-clock-diag-1}).
Note the envelope of the spikes seen in the insulating state has the same shape as the superfluid spectrum.  This can be understood from noting that in both cases the density is proportional to $\mu$ or $\lceil \mu/U\rceil$, resulting in similar coarse-grained $\Delta$.

\subsection{Intermediate Coupling\label{corrug}}

\begin{figure}[hbtp]
\setlength{\unitlength}{1.0in}
\centering
\mbox{
\subfigure[]
{
\begin{picture}(1.6,1.5)
\put(0.,0.){
\includegraphics[width=1.6in,angle=0]{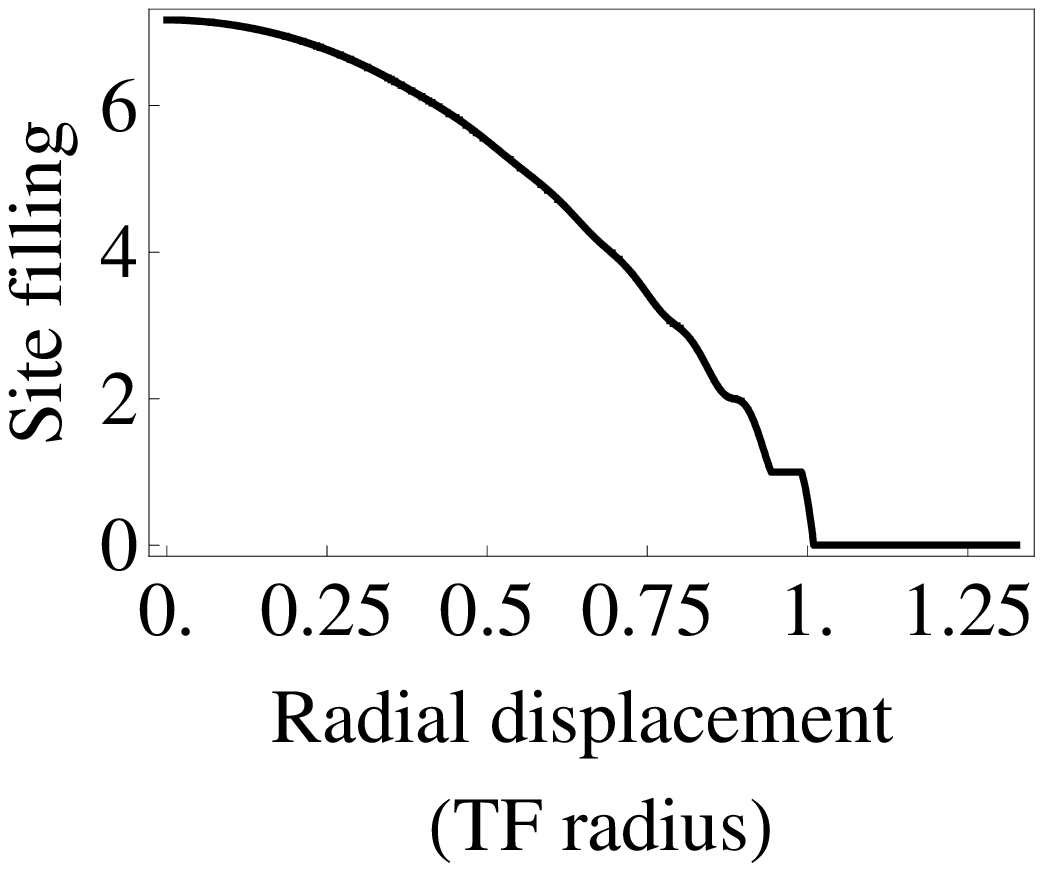}
\label{fig:cs-corrug-realspace}
}
\end{picture}
}
\subfigure[]
{
\begin{picture}(1.6,1.5)
\put(0.,0.2){
\includegraphics[width=1.6in,angle=0]{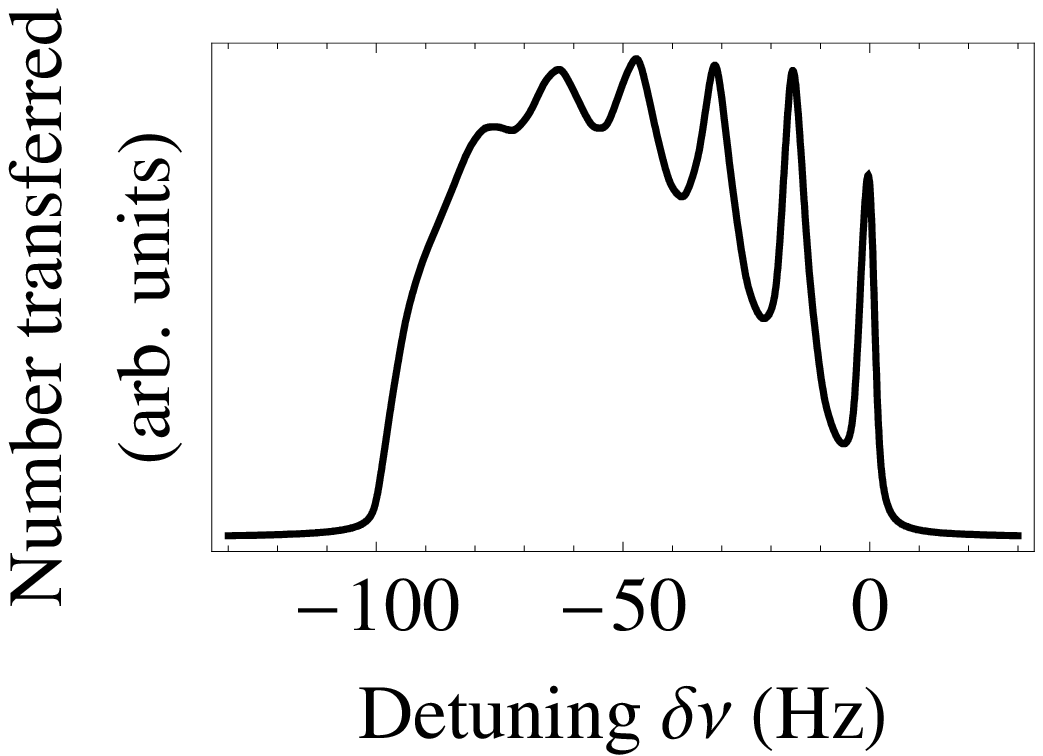}
\label{fig:cs-warning-corrug}
}
\end{picture}
}
}
\caption{ (a)
The density as a function of distance from the trap center for a harmonic trap in units of the Thomas-Fermi radius $\ell_{\text{TF}}$, defined as $\ell_{\text{TF}}\equiv\sqrt{\lp\mu _0-\epsilon _0 \rp/\lp m \left.\omega ^2\right/2\rp}$.
The density profile corresponds to the LDA contour at $t/U_{a}=0.018$ of Fig.~\ref{fig:GMF-LDA-phase-diagram}.  The corrugation of the density is observable, but not dramatic; it would be particularly difficult to image in experiments looking at columnar integrated densities.
(b)
The hyperfine spectra arising from the density plotted in (a).
}
\end{figure}

Finally, let's consider how the spectrum evolves as one increases $t/U$ from zero.  For non-zero $t/U$, superfluid shells form between Mott plateaus.  These regions make the density continuous.  Consequently, in the spectra, the areas of zero signal between peaks begin to fill in.  Using our numerics, we find that
the peaks remain visible until the system is well into the superfluid regime.  An example is shown in Fig.~\ref{fig:cs-warning-corrug}, corresponding to the chemical potential trajectory at $t/U_{a}=0.018$ shown in Fig.~\ref{fig:GMF-LDA-phase-diagram}.   Note that although the only Mott lobe crossed is at $n=1$, six peaks are clearly visible.
 Clearly one must be cautious about using such spectra to distinguish superfluid and Mott insulating states.

The source of the peaks are weak density corrugations which arise in  the  \textit{superfluid}  state near the Mott boundaries.  These corrugations can be inferred from the unequal spacing of the isodensity contours in
 Fig.~\ref{fig:cs-full}.  The spectrum is a powerful amplifier of these corrugations, as they are hardly prominent in the real-space density shown in Fig.~\ref{fig:cs-corrug-realspace}.

\section{Refinements \label{cs-refinements} }

As seen previously, GMF+LDA captures the main features of the experimental spectra: sharp peaks occur in the Mott insulator limit, a smooth distribution in the deep superfluid limit.  Furthermore, the overall energy scales of our spectra are consistent with those found experimentally.  We caution however that we have used identical trapping frequencies in each of our spectra, while experimentally the harmonic confinement varies in an uncharacterized manner when the optical lattice intensity is changed.  With this systematic variation makes quantitative comparison difficult.

Despite the qualitative similarities between theory and experiment,
serious discrepancy remains.  In particular,  the experiment finds much more spectral weight at small detunings than theory predicts.
Here we explore possible sources of this discrepancy.  Our primary result is that the discrepencies are
consistent with
trap anharmonicities.

In Sec.~\ref{anh} we give an analysis of trap anharmonicities.  In the following sections we briefly discuss several other possible explanations of the discrepencies: non-equilibrium effects and nonlinearities in the transfer rate.
Although these latter two effects could distort the spectrum in a manner qualitatively consistent with experiment, we find that neither of  them plays a significant role in these particular experiments.

\subsection{Anharmonicity}\label{anh}

The trap used in the experiments of Ref.~\cite{campbell:ketterle-clock-shift} is a combination of an Ioffe-Pritchard magnetic trap, which is roughly harmonic, and an optical trap, which provide highly anisotropic Gaussian confinement.  This results in a trap with ``soft" anharmonicities, increasing the number of particles in the low density tails of the cloud.  This will accentuate the small $\delta \omega$ peaks in the spectrum.
The presence of anharmonicities is clear in Fig. 4 of Ref.~\cite{campbell:ketterle-clock-shift}, where the spatial distribution of the Mott insulator shells is far from elliptical.

\begin{figure}
\setlength{\unitlength}{1.0in} \centering
\mbox{
\subfigure[] {
\vspace{-1.2in}\includegraphics[width=1.6in,angle=0]{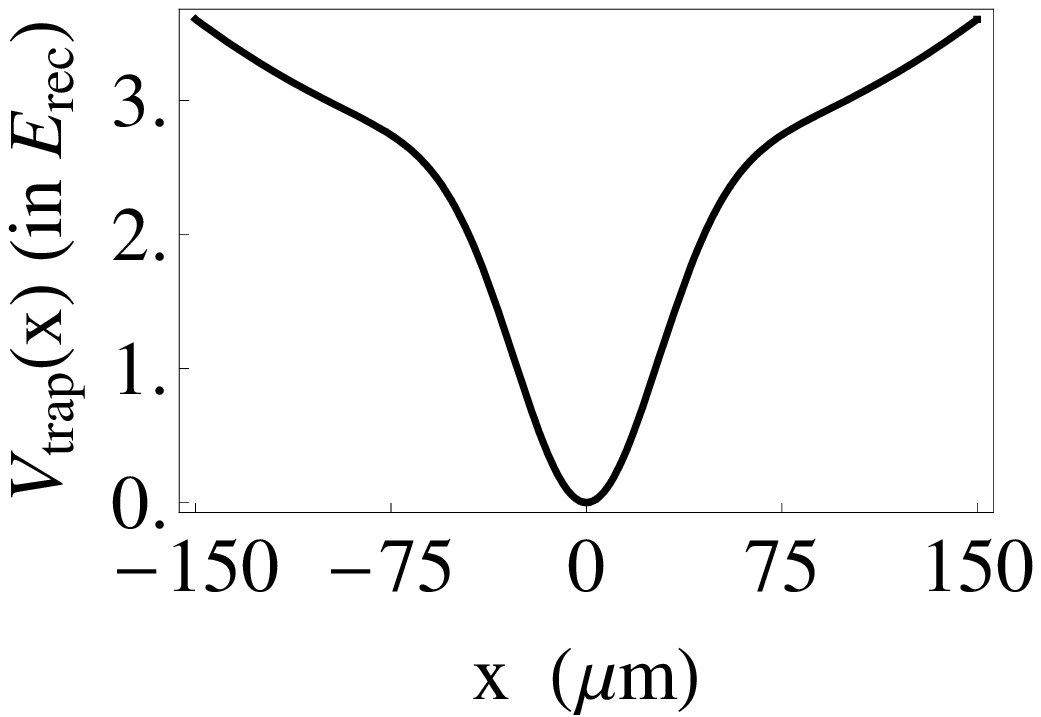}
\label{fig:anh-pot}}
\subfigure[] {
\includegraphics[width=1.6in,angle=0]{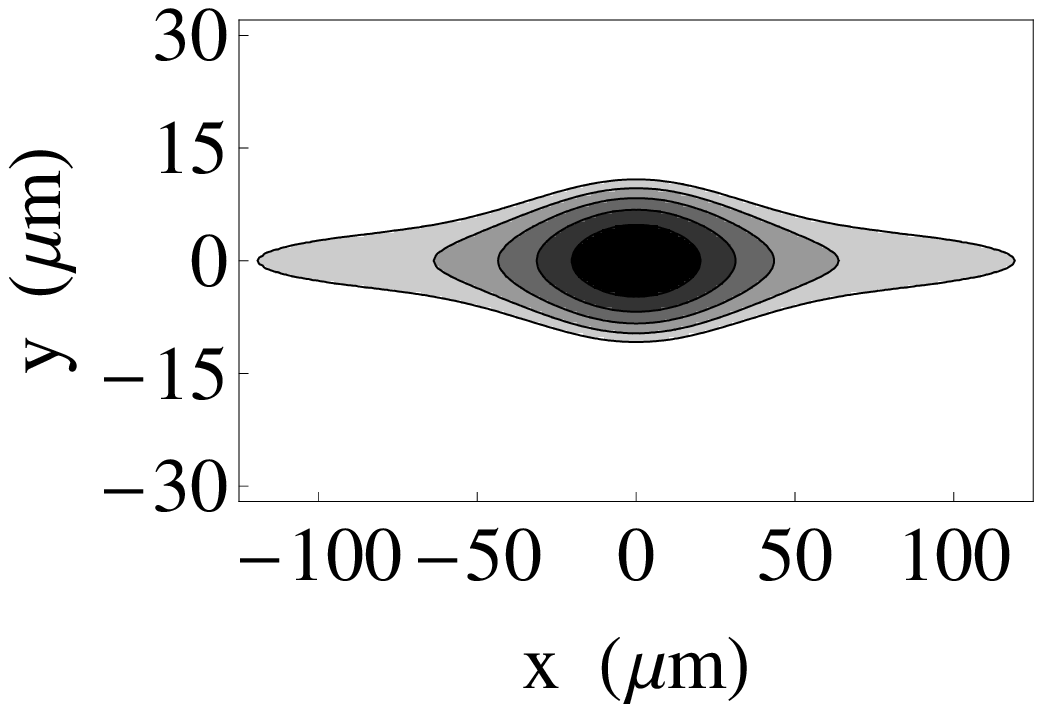}
\label{fig:spatial-anisotropy}}
}
\subfigure[] {
\includegraphics[width=3.25in,angle=0]{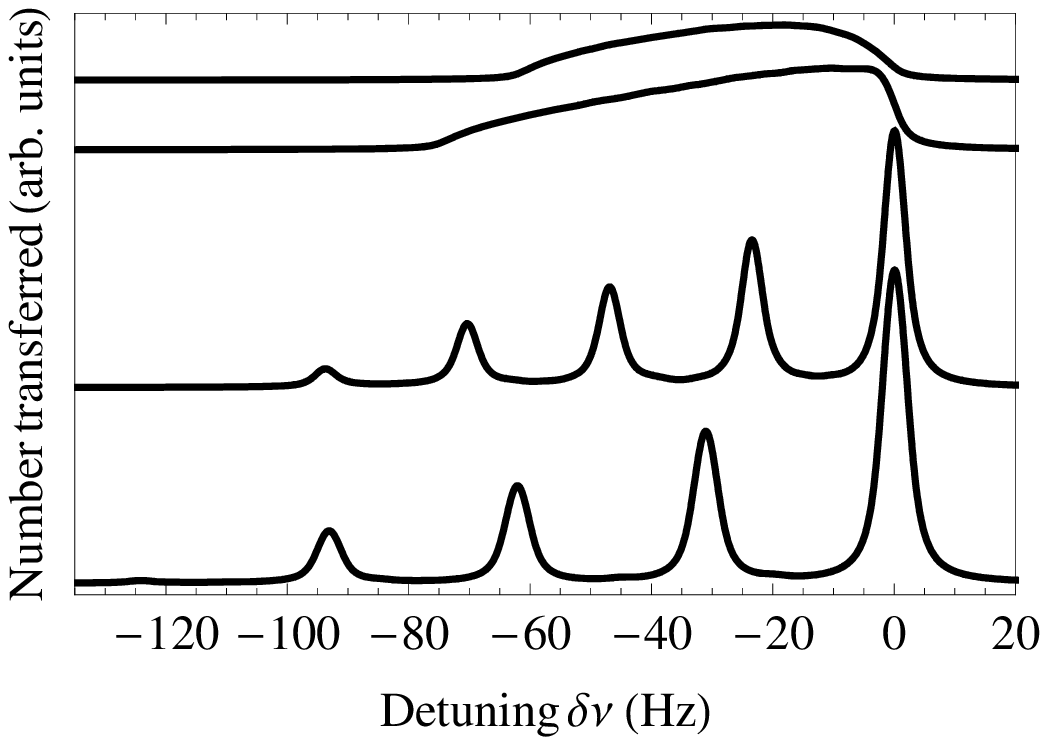}
\label{fig:anh-convolved}
}
\caption{ (a) A slice of the anharmonic potential $V_{\text{trap}}(\v{r})$ similar to the one found in experiments.
(b)
Contour lines of constant density in the x-y plane for $n=1,2,\ldots,5$ in the anharmonic trap   at $V_0=35 E_{\text{rec}}$.
(c)
	Corresponding spectra (using the ``spherical trap model''  discussed in text) for $V_0=5,$ 10, 25, $35E_{\text{rec}}$ with $N=1.4\times10^7$ particles.
}
\end{figure}

We model the trapping potential as
\be
V_{\text{trap}}(x,y,z) &=& \frac{m\omega_a^2}{2} x^2+\frac{m\omega_r^2}{2}(y^2+z^2)\nonumber\\
    &&\hspace{-1in}{}+I_a \lp1-e^{-x^2/(2\sigma^2)}\rp+I_r \lp1-e^{-(y^2+z^2)/(2\sigma^2)}\rp    \label{anh-trap}
\ee
where $x$ lies in the soft ``axial" direction while $y$ and $z$ constitute the ``radial" directions.
The $1/e^2$ beam waist is quoted as $70\mu$m, corresponding to $\sigma=35\mu$m, however we find  spatial profiles closer to experiment from the slightly smaller $\sigma=28\mu$m and use this value throughout. The explicit harmonic terms come from the magnetic trap. The anharmonic Gaussian part has two contributions, $I_{o,j}$ from the optical trap and  $\alpha_j V_0$ from the optical lattice inducing a further trapping potential, for some constants $\alpha_j$, with $I_j=I_{o,j}+\alpha_j V_0$.  The parameters $\omega_a$, $\omega_r$, $\alpha_a$, and $\alpha_r$ are determined from $I_{o,a}$, $I_{o,r}$ and the quadratic trap frequencies $\Omega_{j,V_0}$ at $V_0=0$ and $V_0=40 E_{\text{rec}}$ by matching the quadratic terms of Eq.~(\ref{anh-trap}), giving
\beq
\alpha_j &=& \frac{m\sigma^2}{40 E_{\text{rec}}}\lp\Omega_{j,40}^2-\Omega_{j,0}^2\rp,\\
 \omega_j^2 &=& \Omega_{j,0}^2 - \frac{I_{o,j}}{m\sigma^2}.
\eeq
The quadratic trap frequencies $\Omega_{j,V_0}$ are given in Ref.~\cite{campbell:ketterle-clock-shift}  as $\Omega_{r,0}=2\pi\times70$Hz, $\Omega_{r,40}=2\pi\times110$Hz, $\Omega_{a,0}=2\pi\times20$Hz, and $\Omega_{a,40}=2\pi\times30$Hz.  The remaining unknown parameter $I_o$ is chosen to be $I_{o,a}=1.17 E_{\text{rec}}$ so that
the spatial density profile appears similar to that in the experiment.  We take  $I_{o,r}=I_{o,a}$ though $I_{o,r}$ has little effect on the spatial density profiles.  This yields $\omega_a=4.8$Hz, $\omega_r= 67$Hz, $\alpha_a=0.039 E_{\text{rec}}^{-1}$, and $\alpha_r=0.56 E_{\text{rec}}^{-1}$ to completely characterize the trapping potential of Eq.~(\ref{anh-trap}).
Note, that while we have chosen values to  mimic the experimental observations, we have not attempted to produce a quantitative ``fit" to the experimental data.
Fig \ref{fig:spatial-anisotropy} shows the isopotential lines of our model trap.

For numerical efficiency, we produce spectra from a spherically symmetric model with parameters equal to those of the axial direction, which somewhat exaggerates the anharmonic  effects.  As shown in Fig.~\ref{fig:anh-convolved} the small detuning spectral weight is greatly enhanced by the anharmonicity, producing spectra which are consistent with experiments.

\subsection{Alternative explanations of enhanced low-density spectral weight\label{intern-band-struct}}

Here we examine alternative sources of the enhancement of the small detuning spectral weight observed in experiments.

\textbf{Losses.}
First, we explore the possibility that three-body collisions drive the cloud out of equilibrium.  Atoms are removed preferentially from high density sites, ostensibly enhancing the small-detuning spectral weight.
The timescale for decay from the $n=5$ Mott insulator state is 200ms.  A characteristic equilibration time is the trap period, $\sim 10$ms. Given the separation of  timescales it is extremely unlikely that the system is far out of equilibrium.
Furthermore, the loss rate is effectively zero for one- and two-particle site fillings and hence losses are unable to explain the experimentally observed enhancement of the $n=1$ peak relative to the $n=2$ peak.

\textbf{Nonlinearities in transfer rate.}
The probes used to measure the spectrum may possibly drive the system out of the linear regime where the transfer rate is proportional to the density.   For example, if the transition becomes saturated in the high density regions of the cloud, then the observed spectral weight will be reduced at large detunings.
However, the density dependence of these saturation effects is slow, making it unlikely that they could not be responsible for the dramatic suppression of the ratio of the spectral weight in the $n=2$ and $n=1$ peaks.  A model calculation in the deep Mott regime, where the sites decouple, confirms this result.

\section{Summary}

We calculate the hyperfine spectra of trapped bosonic atoms in an optical lattice.  We consider the cases of harmonic and model anharmonic traps.  We show that a harmonic trap produces a spectrum which shares qualitative features with the experimental spectra measured by Campbell {\em et al.} \cite{campbell:ketterle-clock-shift}: in the deep superfluid limit one has a smooth peak, while in the deep Mott limit, one sees several discrete peaks.
To reproduce the small-detuning spectral weights, however, trap anharmonicities are necessary.

We find the spectra are extremely sensitive to density corrugation. As an example, the mild density corrugations which are found in the superfluid near the Mott insulator boundary are sufficient to produce a pronounced splitting of the spectral line.  Consequently, the spectra are continuous across the superfluid to Mott insulator transition.  Such continuity is characteristic of a second-order phase transition, and makes identifying the superfluid transition difficult.

We acknowledge an illuminating discussion with Jim Sethna regarding the Mott insulating state's clock shift energy.  We thank Wolfgang Ketterle and Gretchen Campbell for information regarding their experiment.  This work was supported  by NSF grant No. PHY-0456261 and GAANN Award No. P200A030111 from the US Department of Education.

\end{document}